\documentclass[bibyear]{aa} 

\usepackage{graphicx}
\usepackage{txfonts}
\begin{document}

   \title{J1342+0928 Supports the Timeline in the $R_{\rm h}=ct$ Cosmology}

   \titlerunning{Premature Quasar Formation}

   \subtitle{}

   \author{F. Melia}

   \institute{Department of Physics, The Applied Math Program, and Department of Astronomy,
The University of Arizona, AZ 85721, USA}

   \date{}

 
  \abstract
   {}
   {The discovery of quasar J1342+0928 ($z=7.54$) reinforces the time
compression problem associated with the premature formation of structure
in $\Lambda$CDM. Adopting the {\it Planck} parameters, we see this quasar
barely 690 Myr after the big bang, no more than several hundred Myr after
the transition from Pop III to Pop II star formation. Yet conventional
astrophysics would tell us that a $10\;M_\odot$ seed, created by a Pop
II/III supernova, should have taken at least 820 Myr to grow via
Eddington-limited accretion. This failure by $\Lambda$CDM constitutes
one of its most serious challenges, requiring exotic `fixes', such
as anomalously high accretion rates, or the creation of enormously massive
($\sim 10^5\;M_\odot$) seeds, neither of which is ever seen in the local
Universe, or anywhere else for that matter. Indeed, to emphasize this point,
J1342+0928 is seen to be accreting at about the Eddington rate, negating
any attempt at explaining its unusually high mass due to such exotic
means. In this paper, we aim to demonstrate that the discovery of 
this quasar instead strongly confirms the cosmological timeline predicted 
by the $R_{\rm h}=ct$ universe.}
   {We assume conventional Eddington-limited accretion and the time versus 
redshift relation in this model to calculate when a seed needed to start
growing as a function of its mass in order to reach the observed mass of
J1342+0928 at $z=7.54$.}
   {Contrary to the tension created in the standard model by the appearance
of this massive quasar so early in its history, we find that in the $R_{\rm h}=ct$
cosmology, a $10\;M_\odot$ seed at $z\sim 15$ (the start of the Epoch of Reionization 
at $t\sim 878$ Myr) would have easily grown into an $8\times 10^8\;M_\odot$ black 
hole at $z=7.54$ ($t\sim 1.65$ Gyr) via conventional Eddington-limited accretion.}
   {}


   \maketitle
%

\section{Introduction}
The recent discovery of ULAS J134208.10+092838.61 (henceforth J1342+0928)
(Banados et al. 2017), an ultraluminous quasar at redshift $z=7.54$, emphasizes
more than ever the time compression problem in the early $\Lambda$CDM universe.
Weighing in at a mass of $M=7.8^{+3.3}_{-1.9}\times 10^8\;M_\odot$, this
supermassive black hole should have taken over 820 Myr to grow via standard
Eddington-limited accretion. Yet we see it barely several hundred Myr
after Pop II and III supernovae could have created the
$\sim 5-25\;M_\odot$ seeds to initiate the black-hole growth. Worse, this
timeline would suggest that J1342+0928 started growing $\sim 130$ Myr {\it before}
the big bang, which is completely unrealistic (Melia 2013a; Melia \& McClintock 2015).
And what is particularly challenging to the concordance model is that
J1342+0928 is seen to be accreting at $1.5^{+0.5}_{-0.4}$ times the
Eddington rate, arguing against any attempt to mitigate the compression
problem by invoking exotic, greatly super-Eddington growth (Volonteri \&
Rees 2005; Pacucci et al. 2015; Inayoshi et al. 2016).

This discovery follows on the heels of another problematic source,
SDSS J010013.02+280225.8, an ultraluminous quasar at $z=6.30$ (Wu
et al. 2015), and about $50$ others uncovered at redshifts
$z>6$ (Fan et al. 2003; Jiang et al. 2007, 2008; Willott et al. 2007, 2010a,
2010b; Mortlock et al. 2011; Venemans et al. 2013; Banados et al. 2014), all
of which contain a black hole with mass $\sim 10^9\;M_\odot$, and all
of which are difficult to accommodate within the standard model's predicted
timeline. Attempts to resolve the mystery of how such large aggregates of matter
could have assembled so quickly in $\Lambda$CDM have generally fallen into
two categories of exotic mechanisms: either an anomalously high accretion
rate (Volonteri \& Rees 2005; Pacucci et al. 2015; Inayoshi et al. 2016),
and/or the creation of enormously massive seeds (Yoo \& Miralda-Escud\'e
2004; Latif et al. 2013; Alexander et al. 2014). But neither
of these is entirely satisfying because no compelling evidence in 
support of such extreme conditions has yet been found. Note, for example,
that J1342+0928 itself is accreting right at the Eddington rate. And 
for other high-$z$ supermassive black holes with a reasonably estimated 
mass, the inferred luminosity has thus far been at, or close to, the
Eddington value (see, e.g., figure~5 in Willott et al. 2010a).

The formation of massive seeds, which in this context implies the birth
of black holes with a mass $\sim 10^5\;M_\odot$, is even more difficult
to confirm observationally. Such events would presumably last too short
a time to offer any meaningful probability of being seen directly. The
best hope would be to find such objects, known as ``intermediate-mass" 
black holes, after they have formed sufficiently nearby for us to be
able to detect their relatively feeble emission. But even here the evidence 
is sparse and inconclusive. A handful of low-luminosity active galactic
nuclei may be such candidates. For example, NGC~4395 at 4 Mpc appears 
to harbor a $\sim 3.6\times 10^5\;M_\odot$ black hole in its center
(Peterson et al. 2005). Some ultra-luminous X-ray sources (ULX's) in 
nearby galaxies may be intermediate-mass black holes with a mass up to
$\sim 1,000\;M_\odot$ (Maccarone et al. 2007), but even these masses
are well below what is required. Some intermediate-mass black holes
may have been seen in globular clusters, e.g., M31 G1, based on the 
stellar velocities measured near their center, but none has yet stood 
up to followup scrutiny (see, e.g., Baumgardt et al. 2003). Most
recently, we have witnessed the LIGO discovery of $\sim 30-50\;M_\odot$
black holes via the gravitational waves they emit as they spiral 
towards an eventual merger in binaries (Abbott et al. 2017). This
opens up the possibility of eventually discovering even more massive
objects during similar merger events, but none have been seen thus
far. It is safe to conclude that massive seeds may be contemplated
theoretically, but no compelling evidence has yet been found to
confirm their existence beyond a possible handful designated as 
dwarf active galactic nuclei. The ambiguity with the latter is,
of course, that these objcts may have simply grown to their observed
intermediate mass via steady accretion rather than having appeared
via some catastrophic event. 

The purpose of this paper is to demonstrate that such
mysterious, unseen processes are not needed to explain the formation
of these supermassive black holes, arguing that the anomaly is not
with the astrophysics, but with the cosmology itself. As we shall
see, the timeline implied by J1342+0928 may be a significant problem
for $\Lambda$CDM, but not at all for the $R_{\rm h}=ct$ universe
(Melia 2007; Melia \& Shevchuk 2012), a Friedmann-Robertson-Walker
(FRW) cosmology with zero active mass (Melia 2016, 2017a). In this
cosmology, a $\sim 10\;M_\odot$ seed created at $z\sim 15-16$, i.e.,
the beginning of the Epoch of Reionization (EoR), would have
grown via conventional Eddington-limited accretion to a mass of
$\sim 8\times 10^8\;M_\odot$ at $z=7.54$, exactly matching the
observed properties of J1342+0928.

\section{The Early Universe}
In the context of $\Lambda$CDM, with {\it Planck} parameters 
$\Omega_{\rm m}=0.307$, $k=0$, $w_\Lambda=-1$ and Hubble constant 
$H_0=67.7$ km s$^{-1}$ Mpc$^{-1}$ (Planck Collaboration 2016),
the Universe is believed to have become transparent at 
$t^{\Lambda{\rm CDM}}\sim 0.4$ Myr, initiating the so-called Dark 
Ages that lasted until the first (Pop III) stars formed several 
hundred Myr later. Reionization presumably started when these 
objects---and subsequently the black holes they spawned---started 
emitting UV radiation, a process that apparently lasted from 
$z\sim 15$ to $z\sim 6$ (Zaroubi 2012; Jiang et al. 2006). The
EoR in the standard model therefore stretched over a cosmic time 
$t^{\Lambda{\rm CDM}}\sim 400-900$ Myr. By comparison, the 
redshift-time relation in $R_{\rm h}=ct$ is given by the relation
\begin{equation}
1+z = {t_0\over t}\;,
\end{equation}
where $t_0=H_0^{-1}$ is the age of the Universe today, in terms 
of the Hubble constant $H_0$. This equation is straightforward 
to derive, noting that $1+z=a(t_0)/a(t)$ in terms of
the expansion factor $a(t)$ (e.g., Weinberg 1972), while
$a(t)=t/t_0$ in the $R_{\rm h}=ct$ universe (Melia \&
Shevchuk 2012). Thus, if we simply
adopt the same {\it Planck} measured value $H_0=67.7$ km s$^{-1}$
Mpc$^{-1}$, the Dark Ages in this cosmology ended at 
$t^{R_{\rm h}=ct}\sim 878$ Myr, while the EoR extended from 
$t^{R_{\rm h}=ct}\sim 878$ Myr to $\sim 2$ Gyr (see fig.~1). 
Note that the redshift range over which reionization
took place is inferred from observations, and is therefore
independent of the cosmology. But each model predicts its
own unique mapping of redshift to age. Thus, although
the EoR lasted from $z\sim 15$ to $z\sim 6$ in both 
cosmologies, the starting and ending times are different. 
With a redshift $z=7.54$, J1342+0928 is being viewed at
cosmic time $t^{R_{\rm h}=ct}\sim 1.65$ Gyr in the $R_{\rm h}=ct$
universe, approximately $772$ Myr after the onset of the EoR, when 
the ramp-up in stellar formation and supernova activity is believed 
to have occurred.

Though not yet fully confirmed, this temporal sequence of
events and epochs in the early Universe is suggested by
many detailed simulations carried out in recent years (Barkana \&
Loeb 2001; Miralda-Escud\'e 2003; Bromm \& Larson 2004; Ciardi
\& Ferrara 2005; Glover 2005; Greif et al. 2007; Wise \& Abel 2008;
Salvaterra et al. 2011; Greif et al. 2012; Jaacks et al. 2012; see
also the recent reviews by Bromm et al. 2009 and Yoshida et al. 2012).
In this scenario, Pop III stars started forming by $z\sim 20$ at the
core of mini halos with mass $\sim 10^6\;M_\odot$ (Haiman et al. 1996;
Tegmark et al. 1997; Abel et al. 2002; Bromm et al. 2002). In
{\it Planck} $\Lambda$CDM, this redshift corresponds to a cosmic
time $t^{\Lambda{\rm CDM}}\sim 200$ Myr. By comparison, Pop III stars 
in $R_{\rm h}=ct$ would have started forming by $z\sim 70$.

This delay of $\sim 200$ Myr between the big bang and the appearance
of the first stars is difficult to circumvent due to the inefficient
cooling of the primordial gas. There was another delay of at least
$\sim 100$ Myr (Yoshida et al. 2004; Johnson et al. 2007)
before Pop II stars could form, while the hot gas expelled by
the Pop III stars cooled and re-collapsed. Thus, black-hole
seeds created during supernova explosions of {\em evolved} Pop II
and III stars would have started their growth more than $\sim
300$ Myr after the big bang, which would not have afforded
them anywhere near enough time to reach $\sim 10^9\;M_\odot$
status by $z\sim 7$ in standard cosmology. Of course, this is
the primary reason proponents of the massive seed scenario
require exotic mechanisms to create $\sim 10^5\;M_\odot$
black holes by other means (Yoo \& Miralda-Escud\'e 2004;
Latif et al. 2013; Alexander et al. 2014).

In conventional astrophysics, the subsequent growth of black-hole
seeds (massive or otherwise) would have been constrained by the
maximum luminosity attainable with the outward radiation pressure
acting on ionized matter under the influence of gravity. In
hydrogen-rich plasma, this limiting power is known as the
Eddington limit $L_{\rm Edd}\approx 1.3\times 10^{38}(M/M_\odot)$
ergs s$^{-1}$. One also needs to know the efficiency $\epsilon$
for converting rest-mass energy into radiation in order to estimate
the accretion rate $\dot{M}$, in which case one then assumes that
$\dot{M}=L_{\rm bol}/\epsilon c^2$, where $L_{\rm bol}$ is the
bolometric luminosity. To allow for all possible variations
of basic accretion-disk theory, one typically adopts a fiducial
value $\epsilon=0.1$ for this quantity (see, e.g., Melia 2009).
Therefore, with Eddington-limited accretion, one may combine
the expressions for $L_{\rm bol}=L_{\rm Edd}$ and $\dot{M}$, i.e., 
\begin{equation}
{dM\over dt}={1.3\times 10^{38}\;{\rm ergs/s}\over\epsilon c^2M_\odot}\;M
\end{equation}
(Salpeter 1964; see also Melia 2013a), whose straightforward solution 
is the so-called Salpeter relation,
\begin{equation}
M(t) = M_{\rm seed}\exp\left({t-t_{\rm seed}\over 45\;{\rm Myr}}\right),
\end{equation}
where $M_{\rm seed}$ ($\sim 5-25\;M_\odot$) is the seed mass produced
at time $t_{\rm seed}$. According to this expression, it would
have taken J1342+0928 approximately $820$ Myr to grow from an initial
black-hole seed of $10\;M_\odot$.

In principle, this growth time could have been shortened by mergers
in the early Universe (Tanaka \& Haiman 2009; Lippai et al. 2009;
Hirschmann et al. 2010). But according to the simulations, 
there are restrictions on how this mechanism could have worked
that mitigate its likelihood of success. On the plus side, detailed 
merger simulations show that the black-hole population always converges 
towards a Gaussian distribution, regardless of the initial seed profile. 
There is therefore some flexibility in the modeling. To comply with
all of the available data, however, $\sim 100\;M_\odot$ seeds would
have had to start forming by $z\sim 40$ (e.g., Tanaka \& Haiman 2009).
This is well before the EoR (which apparently started
at $z\sim 15$). In addition, this creation of seeds could not have
continued after $z\sim 20-30$. The simulations show that if they
did form past this redshift, then there would have been an overproduction
of the mass density in lower-mass (a few $\times 10^5\;M_\odot$ to a 
few $\times 10^7\;M_\odot$) black holes, compared to what is actually
seen (see, e.g., figs.~5 and 6 in Tanaka \& Haiman 2009). In fact, 
without this cutoff, the lower mass black holes would have been 
overproduced by a factor of as much as $100$ to $1,000$.

So the argument that mergers in the early ($\Lambda$CDM) Universe might 
have played a critical role in forming the supermassive black holes at 
high-$z$ does not sit comfortably with our current interpretation of
Pop III star-formation. Our understanding of why the EoR occurred
at $t\sim 400$ Myr is based on our estimate of the cooling time
required to form this first generation of stars, which corresponded
to a redshift (i.e., $\sim 15$) much smaller than $\sim 40$. And it 
would be difficult to understand why these stars stopped forming below 
$z\sim 20-30$, before the EoR even started. The implication is that
some mechanism other than Pop III supernovae would have been
responsible for creating these massive seeds well before the EoR,
yet this would require new, unknown physics and, even more
importantly, there is currently no observational evidence for
such events occurring prior to $z\sim 15$. 

The viability of this scenario has been further mitigated by 
recent arguments showing that the halo abundance was at least an
order of magnitude smaller than previously thought. Johnson et al.
(2013) have recently carried out large ($4$ Mpc$^3$) high-resolution 
simulations of the formation of halos---and Pop III stars within 
them---in the early universe, self-consistently modeling the subsequent metal
enrichment and the stellar radiation produced by the next generation
of stars (i.e., Pop II). It turns out that Pop III and II stars
formed and evolved co-evally down to a redshift $z\sim 6$. These
simulations showed that the enhanced metal enrichment and the feedback 
radiation---which would have included molecule-dissociating Lyman-Werner 
photons responsible for the destruction of the coolants H$_2$ and HD 
required for the condensation of matter in the early Universe---would 
have significantly changed the rate at which halos and Pop III stars formed. 

Specifically, Johnson et al. (2013) found that the Lyman-Werner radiation produced
both near the halos and over cosmological distances would have effectively 
reduced the halo and Pop III star formation rate at $z\gtrsim 10$ by as much
as an order of magnitude compared to previous simulations in which this
radiation was ignored, to a rate per comoving volume of $\sim 10^{-4}\;
M_\odot$ yr$^{-1}$ Mpc$^{-3}$. Ironically, these same effects would have
actually resulted in a higher stellar mass per unit volume by $z\sim 6$ because,
though they negatively impacted the rate of halo and Pop~III star formation,
they extended the time over which Pop III and Pop II formed and evolved
co-evally. In fact, the Pop III star formation rate at $z\sim 6$ is
found to be $\sim 10^{-5}\;M_\odot$ yr$^{-1}$ Mpc$^{-3}$, just an
order of magnitude lower than its peak at $z\sim 10$. But insofar
as the production of halos for mergers in the early Universe is
concerned, this net shift in the time when they would have formed
reduces the volume density of Pop III supernovae---and therefore
the density of black-hole seeds---at a time (corresponding to
$z\gtrsim 10$) when the frequency of collisions and mergers
among these objects would have mattered most to rapidly grow 
the black-hole mass to allow J1342+0928 to appear at $z=7.54$.
 
The bottom line
is that any attempt at explaining the mysterious appearance of
billion-solar mass black holes at $z\sim 7$ in the context of
$\Lambda$CDM faces a very daunting task that is unlikely to
get easier as more of these objects are found at progressively
higher redshifts.

\section{J1342+0928 in $R_{\rm h}=ct$}
Over the past decade, the predictions of $R_{\rm h}=ct$ have
been compared with those of $\Lambda$CDM using over 20 different
kinds of data, from low to high redshifts, and a wide assortment
of observational signatures, including the redshift-time relation,
the redshift dependence of the Hubble constant $H(z)$, and
various distance measures, such as the luminosity and angular-diameter
distances. A summary of these comparative studies and their outcomes
appears in Table~1 of Melia (2017b). In each and every comparison,
$R_{\rm h}=ct$ has been favoured by the data over $\Lambda$CDM.
In other words, there is now compelling evidence to suggest that
a resolution of the time-compression problem associated with the
premature appearance of massive quasars at $z\sim 6-7$ may be
found in the cosmology itself, rather than unseen, exotic `fixes'
to the formation and growth of supermassive black holes.

\begin{figure}
\vspace{1cm}
\centering
\includegraphics[width=9cm]{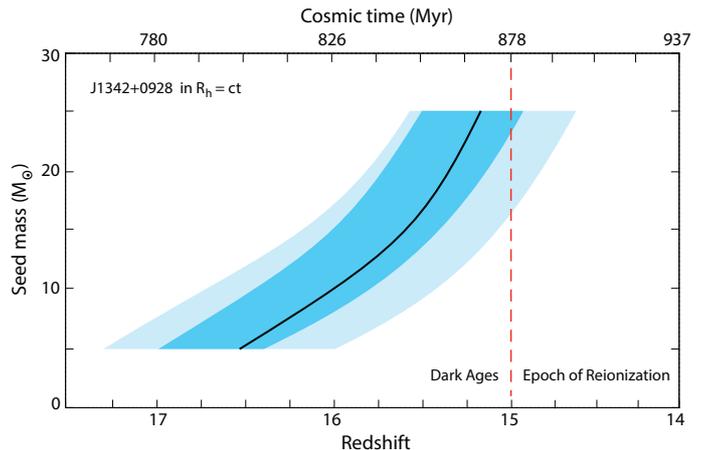}
\caption{Seed mass of J1342+0928 versus redshift at the time it was
formed (solid black curve), assuming this quasar reached its mass
$M=7.8^{+3.3}_{-1.9} \times 10^8\;M_\odot$ at $z=7.54$ via conventional
Eddington-limited accretion. The shaded regions represent the $1\sigma$
(dark) and $2\sigma$ (light) confidence regions based on the $M$ measurement error.}
\end{figure}

In figure~1, we show the seed mass required in $R_{\rm h}=ct$ versus the
time $t_{\rm seed}$ (and corresponding redshift) at which it was produced
in order to account for the appearance of J1342+0928 at $z=7.54$ (solid
black curve). This plot also shows the $1\sigma$ (dark) and $2\sigma$
(light) confidence regions, estimated via error propagation from 
the uncertainty in the measurement of the mass $M$. In other words,
the $1\sigma$ confidence region corresponds to the mass range
$(5.9-11.1)\times 10^8\;M_\odot$.
For comparison, we also see in this figure the demarcation between
the Dark Ages (at $z\gtrsim 15$) and the ensuing EoR ($6\lesssim z
\lesssim 15$). One cannot avoid emphasizing the fact that the
$\sim 820$ Myr required for J1342+0928 to grow via Eddington-limited
accretion from its initial supernova produced $10\;M_\odot$ seed at
$z\lesssim 16$ to its observed $7.8\times 10^8\;M_\odot$ mass at
$z=7.54$ coincides very nicely with two critically important
observations: (1) the redshift range of the EoR, which was apparently
sustained by UV photons emitted by Pop II and III stars, and the quasars
they subsequently spawned; and (2) the approximately Eddington-limited
luminosity observed from J1342+0928 at $z=7.54$.

\section{Discussion and Conclusions}
All the estimates we have made in this paper are based on the assumption 
that high-$z$ quasars accreted steadily at the Eddington rate. We do not know 
their duty cycle, however, so their average growth rate could have been less 
than Eddington. But this just makes the situation worse for the standard model 
because the implied efficiency $\epsilon$ in Equation~(2) would then
be larger in order to achieve the observed final high masses. And
since the characteristic (Salpeter 1964) time ($\tau_{\rm Sal}\sim 45$ Myr) 
scales linearly with $\epsilon$, a bigger efficiency would imply a longer
characteristic time (i.e., $\tau_{\rm Sal}\sim 45 [\epsilon/0.1])$ Myr) in 
the exponential of Equation~(3). In other words, a greater efficiency 
would imply that the same amount of light could be produced with a 
lower mass accretion rate, which would have delayed the growth and 
therefore worsenend the time compression problem. In fact, several 
observations suggest that, when they turned on,
high-$z$ quasars must have accreted at close to Eddington. For example,
Shankar et al. (2009) have argued that only a few $10^{10}\;M_\odot$
black holes have been seen in the local Universe, in spite of
peak quasar activity at $1\lesssim z\lesssim 3$ (McConnell et al. 2012).
Yet quasar masses at $z>3$ would have had to exceed $10^{10}\;M_\odot$
for us to detect their fluxes at Earth if they were sub-Eddington.

On the flip side, one can see from Equation~(3) that J1342+0928
could have grown to its observed mass in only $\sim 270$ Myr
if it had been accreting steadily throughout its growth at $3$ times
the Eddington rate. This would accommodate the timeline in
$\Lambda$CDM, starting with the creation of a $10\;M_\odot$ seed at
$z\sim 15$ growing to $7.8\times 10^8\;M_\odot$ by $z=7.54$.
A similar solution would work for all the other high-$z$ supermassive
black holes as well. But we should then be able to detect at least some
super-Eddington quasars at $z\gtrsim 6$. Unfortunately, all the current
observations rule out such sources (Mortlock et al. 2011; De Rosa et al.
2011; Willott et al. 2010). All the measured accretion rates are at,
or below, the standard Eddington value, with a clear trend towards
even lower rates towards smaller redshifts.

In this paper, we have highlighted the time compression
problem associated with the early appearance of J1342+0928 and
other supermassive black holes at $z> 6$. But today the reality
is that the timeline predicted by $\Lambda$CDM is in conflict with
several kinds of observation, not just the high-$z$ quasars.
The fact that galaxies started forming at $z\sim 10-12$ is
just as difficult to understand (Melia 2014). For example,
with a photometric redshift of $z\approx 10.7$, MACS0647-JD
is the most distant galaxy known reliably to date (Coe et al. 
2013). Its mass is estimated from the typical star-formation rate
measured at lower redshifts and from the inference that the 
average stellar mass ($\sim 10^9\;M_\odot$) of galaxies at 
$z\sim 7-8$ grew to $\sim 10^{10}\;M_\odot$ by $z\sim 2$ 
(Gonzalez et al. 2010). This trend suggests that galaxies
at $z\sim 11$, including MACS0647-JD, have an average stellar 
mass $\la 10^9\;M_\odot$. The problem in $\Lambda$CDM is that
this redshift corresponds to a cosmic time $t\sim 427$ Myr,
implying that about a billion solar masses had to assemble 
inside a galaxy at this redshift in only $\sim 130$ Myr following 
the transition from Pop III to Pop II star formation, which is 
difficult to understand theoretically. Whereas exotic mechanisms 
for the formation and growth of black holes may still be 
considered, there are no such unconventional mechanisms possible 
for creating galaxies.

A diverse set of simulations carried out by independent
workers essentially confirm each other's conclusions because,
in the end, they incorporate the same basic physics. Take
the calculations by Salvaterra et al. (2013) as an illustrative
example that captures the key results. According to their
calculations, the ratio between the mass doubling time 
$t_{\rm db}$ and the cosmic time in these early galaxies is 
universally equal to $\sim 0.1-0.3$, more or less independently 
of redshift. This result appears to be consistent with 
$\sim 10^6-10^8\;M_\odot$ galaxies observed at $z\sim 6-10$,
particularly their measured specific star-formation rate of
$\sim 3-10\;M_\odot$ Gyr$^{-1}$. One can easily show 
(see, e.g., Melia 2014) that
a ratio $t_{\rm db}/t\sim 0.1-0.3$ is sufficient 
to form such galaxies starting with a condensation of
$\sim 10^4\;M_\odot$ at $t\sim 230$ Myr, roughly where one
would expect the transition from Population III
to Population II stars to occur. So there is no problem
forming $10^8\;M_\odot$ galaxies by $z=6$. By the same
token, a $\sim 10^9\;M_\odot$ galaxy at $z=10.7$ (i.e., $t\approx 
490$ Myr in $\Lambda$CDM) would have needed to start growing
at $t\sim 82$ Myr, well before even Pop III stars could have
emerged and exploded, producing the necessary conditions to
begin the subsequent growth of galactic structure. This is
inconsistent with what is thought to have occurred prior
to the end of the dark ages at $t\sim 400$ Myr.

The time compression problem with J1342+0928 in the standard
model therefore has much in common with other evidence suggesting
that the timeline prior to $z\sim 6$ is too short in $\Lambda$CDM.
Taken together, all of the evidence thus far suggests that
the growth of J1342+0928 is best understood in the context of
$R_{\rm h}=ct$. As we have seen, its birth, growth and
evolution are fully consistent with the principal timescales
associated with Pop II and III star formation, and the subsequent
EoR. This result has significant implications because it relies
on the time-redshift relation, rather than integrated distances,
during that crucial early period ($t\lesssim 1-2$ Gyr) of expansion
when cosmologies differ significantly in their respective
predictions. Ultimately, if $R_{\rm h}=ct$ survives as the
correct cosmology, it would obviate the need for inflation,
a considerable shift in the current paradigm (Melia 2013b).

\vskip 0.4in
\begin{acknowledgements}
I am very grateful to the anonymous referee for his/her
helpful comments and suggestions that have led to a significant
improvement in the presentation of this manuscript. I am also 
grateful to Amherst College for its support through a John Woodruff 
Simpson Lectureship, and to Purple Mountain Observatory in Nanjing,
China, for its hospitality while part of this work was being carried out.
\end{acknowledgements}

\end{document}